\documentclass[a4paper,11pt]{amsart}
\usepackage{amssymb}
\usepackage{amsmath}
\usepackage{ascmac}
\usepackage{comment}
\usepackage{wrapfig}
\usepackage{txfonts}
\newtheorem{rem}{Remark}

\def\o#1{\overline{#1}}
\def\u#1{\underline{#1}}

\def\dfrac#1#2{{\displaystyle\frac{#1}{#2}}}
\def\sfrac#1#2{{#1}/{#2}}

\makeatletter
\@addtoreset{equation}{section}

\makeatother

\title{The Pad\'e interpolation method applied to $q$-Painlev\'e equations}
\author{Hidehito Nagao}
\address{Department of Mathematics, Graduate School of Science, Kobe University, Hyogo 657-8501, Japan}
\email{hnagao@math.kobe-u.ac.jp}

\begin{document}

\maketitle



\noindent
{\bf Abstract.}
We establish interpolation problems related to all the $q$-Painlev\'e equations of types from $E_7^{(1)}$ to $(A_2+A_1)^{(1)}$. By solving those problems, we can derive the evolution equations, the scalar Lax pairs and the determinant formulae of special solutions for the corresponding $q$-Painlev\'e equations. 
\vskip5mm



\tableofcontents 

\section{Introduction}\label{sec:intro}
\subsection{The background of discrete Painlev\'e equations}　\\

\noindent
Discrete Painlev\'e equations are discrete equations which reduce to differential Painlev\'e equations by a suitable limiting process. 
As integrable systems, they are studied from various points of view (see \cite{GNR99} for example).

In Sakai's theory \cite{Sakai01}, the discrete Painlev\'e equations were classified on the basis of rational surfaces connected to extended affine Weyl groups. There exist three types of discrete Painlev\'e equations in the classification: elliptic difference ($e$-), multiplicative difference ($q$-) and additive difference ($d$-). The discrete Painlev\'e equations of $q$-difference type are classified as follows:

{\arraycolsep=1pt
\[
\begin{array}{ccccccccccccccccccc}
&&&&&&&&&&&&&&&{\mathbb Z}\\[-2mm]
&&&&&&&&&&&&&&\nearrow\\
&E_8^{(1)}
&\rightarrow&E_7^{(1)}&\rightarrow&E_6^{(1)}
&\rightarrow &\underset{(\mbox{$q$-$P_{\rm VI}$})}{D_5^{(1)}}&\rightarrow&\underset{(\mbox{$q$-$P_{\rm V}$})}{A_4^{(1)}}
&\rightarrow&\underset{(\mbox{$q$-$P_{\rm IV}$,$q$-$P_{\rm III}$})}{(A_2+A_1)^{(1)}}&\rightarrow&\underset{(\mbox{$q$-$P_{\rm II}$})}{(A_1+A_1^{\prime})^{(1)}}
&\rightarrow&\underset{(\mbox{$q$-$P_{\rm I}$})}{A_1^{(1)}}&\rightarrow&{\mathcal D}_6\\[3mm]
\\
\end{array}
\]
}

\subsection{The background of the Pad\'e method}\subsubsection{What is the Pad\'e method?}　\\

\noindent
The method, which we call the {\it Pad\'e method} in this paper, is a method for giving the Painlev\'e equations, the scalar Lax pairs and the determinant formulae of special solutions simultaneously, by starting from suitable problems of Pad\'e approximation/interpolation. 

There exists a close connection between Painlev\'e equations and Pad\'e approximation/interpolation. The Pad\'e method has been presented by applying Pad\'e approximation to differential Painlev\'e equations of types from $P_{\rm VI}$ to $P_{\rm IV}$ in \cite{Yamada09}.

The Pad\'e method is closely related to the theory of orthogonal polynomials (e.g.,\cite{Clarkson13}\cite{Magnus95}\cite{Van07}\cite{Zhedanov05}). By both approaches, we can obtain Painlev\'e equations, Lax pairs and special solutions. (The theory of orthogonal polynomials is more general and the Pad\'e method is simpler.)

\subsubsection{Previous works on the Pad\'e method}　
\\

\noindent
Concerning all the $q$-Painlev\'e equations of types from $E_7^{(1)}$ to $(A_2+A_1)^{(1)}$ in the above classification, the $q$-Painlev\'e equations, the scalar Lax pairs and the special solutions have been already derived by various methods. References are written in the following Table (\ref{eq:painleve-results}):
\begin{equation}\label{eq:painleve-results}
\begin{tabular}{|c||c|c|c|c|c|}
\hline
&$q$-$E_7^{(1)}$&$q$-$E_6^{(1)}$&$q$-$D_5^{(1)}$&$q$-$A_4^{(1)}$&$q$-$(A_2+A_1)^{(1)}$\\
\hline\hline
$\begin{array}{c}q$-Painlev\'e$ \end{array}$&\cite{GR99}&\cite{RGTT01}&\cite{JS96}&
\cite{KTGR00}&
 \cite{KNY01}\cite{KTGR00}\cite{Sakai01}\\
\hline
Lax pair&\cite{Yamada11}&\cite{Sakai06}&\cite{JS96}& \cite{Murata09}& \cite{Murata09}\\
\hline
Special solutions &\cite{Masuda09}&\cite{Ikawa13}&\cite{Sakai98}&\cite{HK07}&\cite{Nakazono10}\\
\hline
\end{tabular}
\end{equation}
Here, amongst the references in the Table (\ref{eq:painleve-results}), only \cite{Ikawa13} is directly related to the Pad\'e method.

The Pad\'e method for discrete Painlev\'e equations has been applied to the following cases: 
\begin{equation}\label{eq:pade-results}
\begin{tabular}{|c|c|c|c|c|}
\hline
&$e$-$E_8^{(1)}$&$q$-$E_8^{(1)}$&$q$-$E_6^{(1)}$&$q$-$D_5^{(1)}$\\
\hline\hline
&\cite{NTY13}&\cite{Yamada14}&\cite{Ikawa13}&\cite{Ikawa13}\cite{Yoshioka10}\\
\hline
Grid &elliptic & $q$-quadric & $q$ & differential, $q$ \\
\hline
\end{tabular}
\end{equation} 
Here, the case $q$-$D_5^{(1)}$ is studied as both the $q$-grid and the differential grid (i.e., Pad\'e approximation).

\subsection{The purpose and the organization of this paper}　\\

\noindent
The purpose of this paper is to apply the Pad\'e method to all the $q$-Painlev\'e equations of types from $q$-$E_7^{(1)}$ to $q$-$(A_2+A_1)^{(1)}$. Namely, by the Pad\'e method, we derive the evolution equations, the scalar Lax pairs and the determinant formulae of special solutions for the corresponding $q$-Painlev\'e equations.  As the main results, the following items are presented for each type.

\vspace{3mm}
{\bf(a)} Setting of the Pad\'e interpolation problem,

{\bf(b)} Contiguity relations,

{\bf(c)} The $q$-Painlev\'e equation,

{\bf(d)} The Lax pair,

{\bf(e)} Special solutions.
\\

\noindent
This paper is organized as follows: In Section \ref{sec:interpolation method}, we will explain the Pad\'e interpolation method applied to the $q$-Painlev\'e equations. Namely, we will explain the methods for deriving items (a)--(e). In Section \ref{sec:interpolation results}, we will present these main results for all the $q$-Painlev\'e equations of types from $E_7^{(1)}$ to $(A_2+A_1)^{(1)}$. In Section \ref{sec:conc}, we will give a summary and discuss some future problems.
\section{Pad\'e interpolation method}\label{sec:interpolation method}　\\
\noindent
In this section, we will explain the methods for deriving the items (a)--(e) in the main results given in Section \ref{sec:interpolation results}.
\subsection{(a) Setting of the Pad\'e interpolation problem}　\\

\noindent
$\bullet$ In this section, we will consider the following interpolation problem (of the $q$-grid):

For a given function $Y(x)$, we look for functions $P_m(x)$ and $Q_n(x)$ which are polynomials of degree $m$ and $n$, satisfying the interpolation condition 
 \begin{equation}\label{eq:q-pade}
Y(q^s)=\sfrac{P_m (q^s)}{Q_n (q^s)} \quad (s=0,1,\ldots, m+n).
\end{equation}
We call this problem the {\it Pad\'e interpolation problem (of the $q$-grid)}. Then, the function $Y(x)$ is called the {\it  interpolated function}. Correspondingly, the polynomials $P_m(x)$ and $Q_n(x)$ are called {\it interpolating functions} respectively.

The common normalization factor of the polynomials $P_m(x)$ and $Q_n(x)$ is not determined by the condition (\ref{eq:q-pade}). However, this normalization factor is not essential for our arguments, i.e., the main results in Section \ref{sec:interpolation results} (see Remark \ref{rem:gauge}). The explicit expressions of $P_m(x)$ and $Q_n(x)$, which will be used in the computations for the item (e) above, were essentially given in \cite{Jacobi} (see item (e) below).\\

\noindent
$\bullet$ In this section, we will establish the interpolation problems (\ref{eq:q-pade}) by specifying the interpolated functions $Y(x)$ and the interpolated sequences $Y_s=Y(q^s)$ as follows:
\begin{equation}\label{eq:Ylist}
\begin{tabular}{|c||c|c|c|c|c|}
\hline
&$q$-$E_7^{(1)}$&$q$-$E_6^{(1)}$&$q$-$D_5^{(1)}$&$q$-$A_4^{(1)}$&$q$-$(A_2+A_1)^{(1)}$\\
\hline\hline
$\begin{array}{c}\\Y(x)\\ \\\end{array}$&$\displaystyle\prod_{i=1}^3\dfrac{(a_ix,b_i;q)_{\infty}}{(a_i,b_ix;q)_{\infty}}$&
$\displaystyle\prod_{i=1}^2\dfrac{(a_ix,b_i;q)_{\infty}}{(a_i,b_ix;q)_{\infty}}$&
$c^{\log_{q}x}\dfrac{(a_1x,b_1;q)_{\infty}}{(a_1,b_1x;q)_{\infty}}$&
$c^{\log_{q}x}\dfrac{(b_1;q)_{\infty}}{(b_1x;q)_{\infty}}$&
$(d\sqrt{\sfrac{x}{q}})^{\log_q x}
$\\
&$\dfrac{a_1a_2a_3q^m}{b_1b_2b_3q^n}=1$&&&&\\
\hline
$\begin{array}{c}\\Y_s\\ \\\end{array}$&$\displaystyle\prod_{i=1}^3\dfrac{(b_i;q)_s}{(a_i;q)_s}$&
$\displaystyle\prod_{i=1}^2\dfrac{(b_i;q)_s}{(a_i;q)_s}$&
$c^s\dfrac{(b_1;q)_s}{(a_1;q)_s}$&
$c^s(b_1;q)_s$&
$q^{\left(\substack{s\\2}\right)}d^s$\\
\hline
$q$-HGF&${}_4\varphi_3$&${}_3\varphi_2$&${}_2\varphi_1$&${}_2\varphi_1$&${}_1\varphi_1$\\
\hline
\end{tabular}
\end{equation}
where $\dfrac{a_1a_2a_3q^m}{b_1b_2b_3q^n}=1$ is a constraint for the parameters in the case $q$-$E_7^{(1)}$, and the $q$-shifted factorials are defined by \begin{equation}\label{eq:qPoch}
\begin{array}{l}
(a_1, a_2, \dots, a_i;q)_j=\displaystyle\prod_{k=0}^{j-1}(1-a_1q^k)(1-a_2q^k)\dots(1-a_iq^k),
\end{array}
\end{equation}
and the $q$-HGF (the $q$-hypergeometric functions \cite{GaR90}) is defined by
\begin{equation}\label{eq:qHGF}
\begin{array}{l}
{}_k\varphi_l\left(
\begin{array}{ccc}
a_1,&\dots,&a_k\\[0mm]
b_1,&\dots,&b_l
\end{array}
; q,x
\right)
=\displaystyle\sum_{s=0}^{\infty}\dfrac{(a_1,\ldots,a_k;q)_s}{(b_1,\ldots,b_l,q;q)_s}\left[(-1)^sq^{\left(\substack{s\\2}\right)}\right]^{1+l-k}x^s,
\end{array}
\end{equation}
with $\left(\substack{s\\2}\right)=\sfrac{s(s-1)}{2}$.

\begin{rem} 
{\rm {\bf On the choice of $Y_s$ and $Y(x)$}

\noindent
In Table (\ref{eq:Ylist}), the expressions for the interpolated sequences $Y_s$ are closely related to those of $q$-HGF ${}_r\varphi_s$. In fact, we chose the $Y_s$ by comparing the coefficient of $x^s$ in ${}_r\varphi_s$ (\ref{eq:qHGF}) and the expression for $Y_s\sfrac{(q^{-(m+n)};q)_s}{(q;q)_s}$ in (\ref{eq:qJacobi}). Then, we also chose the interpolated functions $Y(x)$, which are equal to the $Y_s$ at $x=q^s$. $\square$
}
\end{rem}

\noindent
$\bullet$ In this section, we will consider yet another Pad\'e problem where some parameters $a_i,b_i, c, d, m$ and $n$ in $Y(x)$ are shifted. The parameter shift operators $T$ are given as follows:
\begin{equation}\label{eq:Tlist}
\begin{tabular}{|c||ccc|}
\hline
& Parameters &&\\
\hline\hline
$q$-$E_7^{(1)}$&$(a_1,a_2,a_3,b_1,b_2,b_3,m,n)$&$\mapsto$&$(qa_1,a_2,qa_3,b_1,b_2,qb_3,m-1,n)$\\
\hline
$q$-$E_6^{(1)}$&$(a_1,a_2,b_1,b_2,m,n)$&$\mapsto$&$(qa_1,a_2,b_1,b_2,m-1,n)$\\
\hline
$q$-$D_5^{(1)}$&$(a_1,b_1,c,m,n)$&$\mapsto$&$(qa_1,b_1,c,m-1,n)$\\
\hline
$q$-$A_4^{(1)}$&$(b_1,c,m,n)$&$\mapsto$&$(b_1,c,m-1,n)$\\
\hline
$q$-$(A_2+A_1)^{(1)}$&$(d,m,n)$&$\mapsto$&$(d,m-1,n)$\\
\hline
\end{tabular}
\end{equation}
Here, the operators $T$ are called the {\it time evolutions}, because they specify the directions of the time evolutions for $q$-Painlev\'e equations.\\

\subsection{(b) Computation of contiguity relations}　\\

\noindent
$\bullet$ We will consider the following pair of linear $q$-difference equations $L_2(x)$ and $L_3(x)$ for unknown function $y(x)$, which are the main object in our study. 
\begin{align}\label{eq:L2L3matrix}
L_2(x):
\begin{vmatrix}\nonumber 
y(x) & y(qx) & \o{y}(x) \\
P_m(x) & P_m(qx) & \o{P}_m(x)\\
Y(x)Q_n(x) & Y(qx)Q_n(qx) & \o{Y}(x)\o{Q}_n(x)
\end{vmatrix}=0, \\[3mm]
L_3(x):
\begin{vmatrix} 
y(x) & \o{y}(x) & \o{y}(x/q) \\
P_m(x) & \o{P}_m(x) & \o{P}_m(x/q)\\
Y(x)Q_n(x) & \o{Y}(x)\o{Q}_n(x) & \o{Y}(x/q)\o{Q}_n(x/q)
\end{vmatrix}=0.
\end{align}
Then, the pair of $q$-difference equations (\ref{eq:L2L3matrix}) is called the {\it contiguity relations}. Here, $\o{F}$ and $\u{F}$ denote $T(F)$ and $T^{-1}(F)$ respectively. $T$ is the shift operator acting on parameters given in Table (\ref{eq:Tlist}).\\

\noindent
$\bullet$ We will show the method of computation of the contiguity relations $L_2(x)$ and $L_3(x)$.\\
Set ${\bf y}(x)=\left[\begin{array}{c}P_m(x)\\Y(x)Q_n(x)\end{array}\right]$ and define Casorati determinants $D_i(x)$ by
\begin{equation}\label{eq:Casorati}
\begin{array}{l}
D_1(x)=\det[{\bf y}(x),{\bf y}(qx)],\hspace{3mm} D_2(x)=\det[{\bf y}(x),{\o{\bf y}}(x)],\hspace{3mm} D_3(x)=\det[{\bf y}(qx),\o{{\bf y}}(x)].
\end{array}
\end{equation}
Then, the contiguity relations (\ref{eq:L2L3matrix}) can be rewritten as follows:
\begin{equation}\label{eq:L2L3}
\begin{array}{l}
L_2(x): D_1(x) \o{y}(x)-D_2(x)y(qx)+D_3(x)y(x)=0,
\\
L_3(x): \o{D}_1(\sfrac{x}{q})y(x)+D_3(\sfrac{x}{q}) \o{y}(x)-D_2(x)\o{y}(\sfrac{x}{q})=0.
\end{array}
\end{equation}
We define basic quantities $G(x), K(x)$ and $H(x)$ [e.g., (\ref{eq:E7GKH}), (\ref{eq:E6GKH})] by
\begin{equation}\label{eq:GKH}
\begin{array}{l}
G(x)=\sfrac{Y(qx)}{Y(x)},\quad K(x)=\sfrac{\o{Y}(x)}{Y(x)},\quad H(x)={\rm L.C.M}(G_{\rm den}(x), K_{\rm den}(x)),
\end{array}
\end{equation}
where $G_{\rm den}(x)$ and $G_{\rm num}(x)$ are defined as the polynomials of the denominator and the numerator of $G(x)$ respectively, and $K_{\rm den}(x)$ and $K_{\rm num}(x)$ are similarly defined.
Substituting these quantities into the equations (\ref{eq:Casorati}), we obtain the following determinants:
\begin{equation}\label{eq:Drelation}
\begin{array}{l}
D_1(x)=\dfrac{Y(x)}{G_{\rm den}(x)}\{G_{\rm num}(x)P_{m}(x)Q_n(qx)-G_{\rm den}(x)P_{m}(qx)Q_n(x) \},
\\[5mm]
D_2(x)=\dfrac{Y(x)}{K_{\rm den}(x)}\{K_{\rm num}(x)P_{m}(x)\o{Q}_n(x)-K_{\rm den}(x)\o{P}_{m}(x)Q_n(x) \},
\\[5mm]
D_3(x)=\dfrac{Y(x)}{H(x)}
\{\dfrac{H(x)}{K_{\rm den}(x)}K_{\rm num}(x)P_{m}(qx)\o{Q}_n(x)-\dfrac{H(x)}{G_{\rm den}(x)}G_{\rm num}(x)\o{P}_{m}(x)Q_n(qx)\}.
\end{array}
\end{equation}
Using the interpolation condition (\ref{eq:q-pade}) and the form of the basic quantities [e.g., (\ref{eq:E7GKH}), (\ref{eq:E6GKH})], we can investigate positions of zeros and degrees of the polynomials within braces $\{\hspace{3mm}\}$ of the equations (\ref{eq:Drelation}). Then, we can simply compute the determinants $D_i(x)$ [e.g., (\ref{eq:E7D}), (\ref{eq:E6D})] except for some factors such as $1-fx$ in $D_1(x)$ and $1-\sfrac{x}{g}$ in $D_3(x)$, where $f$ and $g$, etc, are constants with respect to $x$. In this way, we obtain the contiguity relations $L_2(x)$ and $L_3(x)$ [e.g., (\ref{eq:E7L2L3}), (\ref{eq:E6L2L3})].
\begin{rem}\label{rem:gauge}
{\rm {\bf On the gauge invariance of $C_0C_1$}

\noindent
When the common normalization factor of $P_m(x)$ and $Q_n(x)$ is changed, an $x$-independent gauge transformation of $y(x)$ is induced in $L_2(x)$ and $L_3(x)$. Under the $x$-independent gauge transformation of $y(x)$: $y(x)\mapsto Gy(x)$, the coefficients of $\o{y}(x), y(x/q), y(x)$ and $y(x), \o{y}(x), \o{y}(x/q)$ in (\ref{eq:L2L3}) change as follows:
\begin{equation}
\begin{array}{l}
(D_1(x) : D_2(x) : D_3(x))\mapsto(\sfrac{\o{G}D_1(x)}{G} : D_2(x) : D_3(x))\\
(\o{D}_1(x/q) : D_3(x/q) : D_2(x))\mapsto(\sfrac{G\o{D}_1(x/q)}{\o{G}} : D_3(x/q) : D_2(x)).
\end{array}
\end{equation}
The coefficients $C_0$ and $C_1$ in $L_2(x)$ and $L_3(x)$ [e.g., (\ref{eq:E7L2L3}), (\ref{eq:E6L2L3})] are defined as the normalization factors of the coefficients of $\o{y}(x)$ and $y(x)$ respectively. Then, $C_0$ and $C_1$ change under the gauge transformation, but the product $C_0C_1$ is a gauge invariant quantity. Moreover, $C_0$ and $C_1$ do not appear in the final form of the $q$-Painlev\'e equations. $\square$
}
\end{rem}
\subsection{(c) Computation of the $q$-Painlev\'e equation}　\\

\noindent
We can derive the $q$-Painlev\'e equation from the compatibility condition of the contiguity relations $L_2(x)$ and $L_3(x)$ [e.g., (\ref{eq:E7L2L3}), (\ref{eq:E6L2L3})]. Computing the compatibility condition, we determine three quantities $\u{g}, \o{f}$ and $C_0C_1$. Expressions for $\u{g}$ and $\o{f}$ are obtained in terms of $f$ and $g$. An expression for $C_0C_1$ is obtained in terms of $f, g$ and $\o{f}$ (and hence in terms of $f$ and $g$).

The first and the second expressions are the $q$-Painlev\'e equation [e.g., (\ref{eq:E7eq}), (\ref{eq:E6eq})]. The third expression is a constraint for the product $C_0C_1$ [e.g., (\ref{eq:E7C0C1}), (\ref{eq:E6C0C1})].
\begin{rem}\label{rem:generic}
{\rm {\bf On two meanings of $f, g, m$ and $n$}

\noindent
We use the variables $f$ and $g$ with two different meanings. The first meaning is the $f$ and $g$ which are explicitly determined in terms of parameters $a_i, b_i, m$ and $n$ by the Pad\'e problem. The second meaning is the $f$ and $g$ which are unknown functions in the $q$-Painlev\'e equation. In items (c) and (d), we consider $f$ and $g$ in the second meaning.

Similarly, we use the parameters $m$ and $n$ with two meanings. In the first meaning, $m$ and $n$ are integers. In the second meaning, $m$ and $n$ are generic complex parameters. In items (c) and (d), we consider $m$ and $n$ in the second meaning. Then, the result of the compatibility of $L_2(x)$ and $L_3(x)$ also holds with respect to the second meaning. $\square$
}
\end{rem}
\subsection{(d) Computation of the Lax pair}　\\

\noindent
$\bullet$ We will consider the following pair of linear $q$-difference equations for the unknown function $y(x)$:
\begin{equation}\label{eq:L1L2}
\begin{array}{l}
L_1(x): A_1(x)y(\sfrac{x}{q})+A_2(x)y(x)+A_3(x)y(qx)=0,
\\
L_2(x): A_4(x)\o{y}(x)+A_5(x)y(x)+A_6(x)y(qx)=0,
\end{array}
\end{equation}
such that their compatibility condition gives a $q$-Painlev\'e equation. Then, the pair of $q$-difference equations is called the "scalar Lax pair".\\

\noindent
$\bullet$ We will show the method of computation of the scalar Lax pair (\ref{eq:L1L2}). \\
The Lax pair $L_1(x)$ and $L_2(x)$, which satisfies the compatibility condition, is derived using the results of items (a)--(c) as follows: The $L_2(x)$ equation in item (d) [e.g., (\ref{eq:E7L1L2}), (\ref{eq:E6L1L2})] is the same as the $L_2(x)$ in the item (b). We can obtain the Lax equation $L_1(x)$ as follows: First, combining the contiguity relations $L_2(x)$ and $L_3(x)$ [e.g., (\ref{eq:E7L1L2}), (\ref{eq:E6L1L2})], one obtains an equation between the three terms $y(q x), y(x)$ and $y(x/q)$ (See the figure below), whose coefficient functions depend on the variables $f, g, \o{f}, C_0$ and $C_1$. However, the variables $C_0$ and $C_1$ appear through the product $C_0C_1$. Therefore, expressing $\o{f}$ and $C_0C_1$ in terms of $f$ and $g$ only, one obtains the Lax equation $L_1(x)$ [e.g., (\ref{eq:E7L1L2}), (\ref{eq:E6L1L2})].

\begin{center}\setlength{\unitlength}{1.2mm}
\begin{picture}(50,28)(-5,-3)
\put(-1,23){$\o{y}(\sfrac{x}{q})$}
\put(20,23){$\o{y}(x)$}
\put(-1,-4){$y(\sfrac{x}{q})$}
\put(20,-4){$y(x)$}
\put(41,-4){$y(qx)$}
\put(10,-4){$L_1(x)$}
\put(0,-1){\line(1,0){42}}
\put(0,0){\line(1,0){20}}
\put(0,0){\line(0,1){20}}
\put(0,20){\line(1,-1){20}}
\put(1,21){\line(1,0){20}}
\put(21,1){\line(0,1){20}}
\put(1,21){\line(1,-1){20}}
\put(2,4){$L_2(\sfrac{x}{q})$}
\put(11,14){$L_3(x)$}
\put(22,0){\line(1,0){20}}
\put(22,0){\line(0,1){20}}
\put(22,20){\line(1,-1){20}}
\put(26,4){$L_2(x)$}
\end{picture}
\end{center}
\subsection{(e) Computation of special solutions}　\\

\noindent
By construction, expressions for $f$ and $g$ as in the first meaning in Remark \ref{rem:generic} give a special solution for the $q$-Painlev\'e equation.
We will present how to compute determinant formulae of the special solutions.\\

\noindent
$\bullet$ We will derive formulae (\ref{eq:qJacobi}) which are convenient for computing the special solutions $f$ and $g$. The Cauchy--Jacobi formulae (\ref{eq:Jacobi}) are essentially presented in \cite{Jacobi}. For a given sequence $Y_s$, the polynomials $P_m(x)$ and $Q_n(x)$ of degree $m$ and $n$ for an interpolation problem 
\begin{equation} \label{eq:pade}
Y_s=\sfrac{P_m (x_s)}{Q_n (x_s)} \quad (s=0, 1, \dots, m+n)
\end{equation}
are given by the following determinant expressions:
\begin{equation}\label{eq:Jacobi}
P_{m}(x)=F(x)\det\Big[\sum^{m+n} _{s=0}u_s \dfrac{x_s^{i+j}}{x-x_s }\Big]^n _{i,j=0}, \quad Q_n(x)=\det\Big[\sum^{m+n} _{s=0}u_s x_s^{i+j}(x-x_s )\Big]^{n-1} _{i,j=0},
\end{equation}
where $u_s =\sfrac{Y_s}{F^{\prime}(x_s)}$ and $F(x)=\prod_{i=0}^{m+n}(x-x_i )$.

In the $q$-grid case of problem (\ref{eq:pade}) (i.e., the case of problem (\ref{eq:q-pade})), the formulae  (\ref{eq:Jacobi}) take the following form:
\begin{equation}\label{eq:qJacobi}
\begin{array}{l}
P_{m}(x)=\dfrac{F(x)}{(q;q)_{m+n}^{n+1}}\det\Big[\displaystyle\sum^{m+n} _{s=0}Y_s\dfrac{(q^{-(m+n)};q)_s }{(q;q)_s}\dfrac{q^{s(i+j+1)}}{x-q^s }\Big]^n _{i,j=0},
\\
Q_n(x)=\dfrac{1}{(q;q)_{m+n}^{n}}\det\Big[\displaystyle\sum^{m+n} _{s=0}Y_s\dfrac{(q^{-(m+n)};q)_s }{(q;q)_s }q^{s(i+j+1)}(x-q^s)\Big]^{n-1} _{i,j=0}.
\end{array}
\end{equation}

In the derivation of (\ref{eq:qJacobi}), we have used the following relations:
\begin{align}\label{eq:Fprime}
F^{\prime}(x_s )=&(x_s -x_0 )\dots(x_s -x_{s-1})(x_s -x_{s+1})\dots(x_s -x_{m+n})\nonumber \\
=&(-1)^s q^{(s-1)s/2}(q;q)_s q^{s(m+n-s)}(q;q)_{m+n-s}\nonumber\\
=&\sfrac{(q;q)_s (q;q)_{m+n}}{q^{s}(q^{-(m+n)};q)_s}.  
\end{align}
Moreover, substituting the values of $Y_s$ (\ref{eq:Ylist}) and $F^{\prime}(x_s )$ (\ref{eq:Fprime}) into the formulae (\ref{eq:Jacobi}), one obtains the determinant formulae (\ref{eq:qJacobi}).\\

\noindent
$\bullet$ We will show the method of computation of the special solutions $f$ and $g$.\\
The expressions for the $f$ and $g$ can be derived by comparing $D_i(x)$ in (\ref{eq:Drelation}) and $D_i(x)$ [e.g., (\ref{eq:E7D}), (\ref{eq:E6D})] in item (b) as the identity with respect to the variable $x$ and applying the formulae (\ref{eq:qJacobi}). 

For example, the computation for the case $q$-$E_7^{(1)}$ is as follows: Substituting $x=\sfrac{1}{a_i}$ into $D_1(x)$ in (\ref{eq:Drelation}) and $D_1(x)$ in (\ref{eq:E7D}), we obtain an expression for the $f$ in the first equation of (\ref{eq:E7sol}) by comparing the two expressions for $D_1(x)$ and applying the formulae (\ref{eq:qJacobi}). Similarly, substituting $x=\sfrac{1}{b_i}$ into the third equation of (\ref{eq:Drelation}) and $D_3(x)$ in  (\ref{eq:E7D}), we obtain an expression for the $g$ in the second equation of (\ref{eq:E7sol}) by comparing the two expressions for $D_3(x)$ and applying the formulae (\ref{eq:qJacobi}).\\
\section{Main results}\label{sec:interpolation results}　\\

\noindent
In this subsection, we will present the results obtained through the method, which is explained for each case from $q$-$E_7^{(1)}$ to $q$-$(A_2+A_1)^{(1)}$ in subsection \ref{sec:interpolation method}.

We use the following notations:
\begin{equation}\label{eq:notations}
\begin{array}{l}
\sfrac{a_1a_2\dots a_n}{b_1b_2\dots b_n}=\dfrac{a_1a_2\dots a_n}{b_1b_2\dots b_n},\\
{\mathcal N}(x)=\displaystyle\prod_{i=0}^{m+n-1}(1-\sfrac{x}{q^i}),\\
T_{a_i}(F)=F |_{a_i \to qa_i},\quad T_{a_i}^{-1}(F)= F |_{a_i \to \sfrac{a_i}{q}},
\end{array}
\end{equation}
for any quantity (or function) $F$ depending on variables $a_i$ and $b_i$.\\
\subsection{Case $q$-$E_7^{(1)}$}\label{subsec:E7} 　\\

\noindent
{\bf(a)} Setting of the Pad\'e interpolation problem\\

The interpolated function, the interpolated sequence and the constraint (\ref{eq:Ylist}):
\begin{equation}\label{eq:E7Y}
\begin{array}{l}
Y(x)=\displaystyle\prod_{i=1}^3\dfrac{(a_i x, b_i;q)_\infty}{(a_i, b_i x;q)_\infty},
\quad
Y_s=\displaystyle\prod_{i=1}^{3}\dfrac{(b_i;q)_s}{(a_i;q)_s},
\quad \dfrac{a_1a_2a_3q^m}{b_1b_2b_3q^n}=1.
\end{array}
\end{equation}

The time evolution (\ref{eq:Tlist}):
\begin{equation}\label{eq:E7T}
T: (a_1,a_2,a_3,b_1,b_2,b_3,m,n) \mapsto (qa_1,a_2,qa_3,b_1,b_2,qb_3,m-1,n).
\end{equation}\\

\noindent
{\bf(b)} Contiguity relations\\

The basic quantities:
\begin{equation}\label{eq:E7GKH}
G(x)=\displaystyle\prod_{i=1}^{3}\dfrac{(1-b_ix)}{(1-a_ix)},\quad K(x)=\dfrac{1-b_3x}{1-b_3}\displaystyle\prod_{i=1,3}\dfrac{(1-a_i)}{(1-a_ix)},\quad H(x)=(1-b_3)\displaystyle\prod_{i=1}^{3}(1-a_ix).
\end{equation}

The Casorati determinants:
\begin{equation}\label{eq:E7D}
\begin{array}{l}
D_1(x)=\dfrac{c_0x(1-xf){\mathcal N}(x)Y(x)}{G_{\rm den}(x)},\quad
D_2(x)=\dfrac{c_1(1-\sfrac{b_3x}{a_2q^mg}){\mathcal N}(x)Y(x)}{K_{\rm den}(x)},\\
D_3(x)=\dfrac{c_1(1-b_3x)(1-\sfrac{x}{g}){\mathcal N}(x)Y(x)}{H(x)},
\end{array}
\end{equation}
where $f, g, c_0$ and $c_1$ are constants with respect to $x$. 

The contiguity relations:
\begin{equation}\label{eq:E7L2L3}
\begin{array}{l}
L_2(x): C_0x(1-xf)\o{y}(x)-(1-a_2 x)(1-\sfrac{b_3x}{a_2q^mg})y(qx)+(1-b_3x)(1-\sfrac{x}{g})y(x)=0,
\\
L_3(x): C_1x(1-\sfrac{x\o{f}}{q})y(x)+\dfrac{A_2(x)}{1-b_3x}(1-\dfrac{x}{qg})\o{y}(x)-\dfrac{A_1(\sfrac{x}{q})}{1-\sfrac{a_2x}{q}}(1-\sfrac{b_3x}{a_2q^mg})\o{y}(\sfrac{x}{q})=0,
\end{array}
\end{equation}
where
\begin{equation}
\begin{array}{l}
A_1(x)=(1-a_2x)(1-qx)\displaystyle\prod_{i=1,2}(1-b_ix), 
\\
A_2(x)=(1-b_3x)(1-\sfrac{x}{q^{m+n}})\displaystyle\prod_{i=1,3}(1-a_ix),\\
C_0=\sfrac{c_0(1-b_3)}{c_1},\quad C_1=\sfrac{\o{c}_0(1-a_1)(1-a_3)}{qc_1}.
\end{array}
\end{equation}\\

\noindent
{\bf(c)} The $q$-Painlev\'e equation\\

Compatibility gives the following equations:
\begin{equation}\label{eq:E7eq}
\begin{array}{l}
\dfrac{(fg-1)(f\u{g}-1)}{(fg-\sfrac{b_3}{a_2q^m})(f\u{g}-\sfrac{b_3}{a_2q^{m+1}})}=\dfrac{A_1(\sfrac{1}{f})}{A_2(\sfrac{1}{f})},\\
\dfrac{(1-fg)(1-\o{f}g)}{(1-\sfrac{a_2q^mfg}{b_3})(1-\sfrac{a_2q^{m-1}\o{f}g}{b_3})}=\dfrac{A_1(g)}{A_2(\sfrac{a_2q^mg}{b_3})}.
\end{array}
\end{equation}
These equations (\ref{eq:E7eq}) are equivalent to the $q$-Painlev\'e equation of type $E_7^{(1)}$ given in \cite{GR99} \cite{KMNOY04} \cite{RGTT01}.
The 8 singular points are on the two curves $fg=1$ and $fg=\sfrac{b_3}{a_2q^m}$.
\begin{equation}
\begin{array}{l}
(f,g)=(a_2,\sfrac{1}{a_2}),(b_1,\sfrac{1}{b_1}),(b_2,\sfrac{1}{b_2}),(q,\sfrac{1}{q}),\\
\phantom{(f,g)=}(a_1,\sfrac{b_3}{a_1a_2q^m}),(b_3,\sfrac{1}{a_2q^m}),(\sfrac{1}{q^{m+n}},\sfrac{b_3q^n}{a_2}),(a_3,\sfrac{b_3}{a_2a_3q^m}).
\end{array}
\end{equation}
The product $C_0C_1$:
\begin{equation}\label{eq:E7C0C1}
C_0C_1=\dfrac{A_1(g)(1-\sfrac{b_3}{a_2q^m})(1-\sfrac{b_3}{a_2q^{m-1}})}{q(1-fg)(1-\o{f}g)g^2}.
\end{equation}\\

\noindent
{\bf(d)} The Lax pair
\begin{equation}\label{eq:E7L1L2}
\begin{array}{l}
L_1(x): \dfrac{(b_3-a_2q^m)x^2}{a_2b_3q^mg}\left[\dfrac{A_1(g)}{(fg-1)(qg-x)}-\dfrac{A_2(\sfrac{a_2q^mg}{b_3})}{(\sfrac{a_2q^mfg}{b_3}-1)(\sfrac{a_2q^mg}{b_3}-x)}\right]y(x)
\\[5mm]
\phantom{L_1(x): }+\dfrac{q^2(q-b_3x)A_1(\sfrac{x}{q})}{(q-a_2x)(q-fx)}\left[y(\sfrac{x}{q})-\dfrac{b_3(\sfrac{a_2q^{m+1}g}{b_3}-x)(q-a_2x)}{a_2q^m(qg-x)(q-b_3x)}y(x)\right]
\\[5mm]
\phantom{L_1(x): }+\dfrac{(1-a_2x)A_2(x)}{(1-b_3x)(1-fx)}\left[y(qx)-\dfrac{a_2q^m(g-x)(1-b_3x)}{b_3(\sfrac{a_2q^mg}{b_3}-x)(1-a_2x)}y(x)\right]=0,
\\[5mm]
L_2(x): C_0x(1-xf)\o{y}(x)-(1-a_2 x)(1-\sfrac{b_3x}{a_2q^mg})y(qx)
\\[3mm]
\phantom{L_2(x): C_0x(1-xf)\o{y}(x)}+(1-b_3x)(1-\sfrac{x}{g})y(x)=0.
\end{array}
\end{equation}
The scalar Lax pair (\ref{eq:E7L1L2}) is equivalent to that in \cite{Yamada11} using a suitable gauge transformation of $y(x)$. (Note that there is a typographical error in the second equation of (36) in \cite{Yamada11}, namely the expression $\o{f}g-t^2$ should read $\o{f}gq-t^2$.)\\

\noindent
{\bf(e)} Special solutions
\begin{equation}\label{eq:E7sol}
\begin{array}{l}
\dfrac{1-\sfrac{f}{a_1}}{1-\sfrac{f}{a_2}}=\dfrac{\gamma_1 T_{a_1}(\tau_{m,n})T_{a_1}^{-1}(\tau_{m+1,n-1})}{\gamma_2 T_{a_2}(\tau_{m,n})T_{a_2}^{-1}(\tau_{m+1,n-1})},\quad \dfrac{1-\sfrac{1}{b_1g}}{1-\sfrac{1}{b_2g}}=\dfrac{\omega_1 T_{b_1}^{-1}(\tau_{m,n})T_{b_1}(\o{\tau}_{m+1,n-1})}{\omega_2 T_{b_2}^{-1}(\tau_{m,n})T_{b_2}(\o{\tau}_{m+1,n-1})},
\end{array}
\end{equation}
where 
\begin{equation}\label{eq:E7tau}
\tau_{m,n}=\det\Big[{}_4\varphi_3 \Big(\substack{\displaystyle{b_1,b_2,b_3,q^{-(m+n)}}\\[3mm]{\displaystyle{a_1,a_2,a_3}}};q,q^{i+j+1}\Big)\Big]^n _{i,j=0},
\end{equation}
\begin{equation}\label{eq:E7gamma}
\begin{array}{l}
\gamma_i=\dfrac{a_i(1-a_iq^{m+n})(1-\sfrac{a_i}{q})^n\prod_{k=1}^{3}(1-\sfrac{b_k}{a_i})}{(1-a_i)^{n+1}},\quad
 \omega_i=\dfrac{(1-\sfrac{a_2}{b_i})(1-b_i)^n}{(1-\sfrac{b_i}{q})^n},\quad\mbox{for $i=1,2$}.
\end{array}
\end{equation}
These determinant formulae of hypergeometric solutions (\ref{eq:E7sol}) are expressed
in terms of the terminating balanced ${}_4\varphi_3$ (\ref{eq:qHGF}) series (Askey--Wilson polynomials \cite{KMNOY04}), and they are expected to be equivalent to the terminating case of  that in \cite{Masuda09}. \\
\subsection{Case $q$-$E_6^{(1)}$}\label{subsec:E6}　\\

\noindent
The contents of this subsubsection is the same as \cite{Ikawa13}.\\

\noindent
{\bf(a)} Setting of the Pad\'e interpolation problem\\

The interpolated function and the interpolated sequence (\ref{eq:Ylist}):
\begin{equation}\label{eq:E6Y}
\begin{array}{l}
Y(x)=\displaystyle\prod_{i=1}^2\dfrac{(a_i x, b_i;q)_\infty}{(a_i, b_i x;q)_\infty},
\quad
Y_s=\displaystyle\prod_{i=1}^{2}\dfrac{(b_i;q)_s}{(a_i;q)_s}.
\end{array}
\end{equation}

The time evolution (\ref{eq:Tlist}):
\begin{equation}\label{eq:E6T}
T: (a_1,a_2,b_1,b_2,m,n) \mapsto (qa_1,a_2,,b_1,b_2,m-1,n).
\end{equation}\\

\noindent
{\bf(b)} Contiguity relations\\

The basic quantities:
\begin{equation}\label{eq:E6GKH}
G(x)=\displaystyle\prod_{i=1}^{2}\dfrac{(1-b_ix)}{(1-a_ix)},\quad 
K(x)=\dfrac{1-a_1}{1-a_1x},\quad H(x)=\prod_{i=1}^{2}(1-a_ix).
\end{equation}

The Casorati determinants:
\begin{equation}\label{eq:E6D}
\begin{array}{l}
D_1(x)=\dfrac{c_0x(1-xf){\mathcal N}(x)Y(x)}{G_{\rm den}(x)},\quad
D_2(x)=\dfrac{c_1{\mathcal N}(x)Y(x)}{K_{\rm den}(x)},\quad
D_3(x)=\dfrac{c_1(1-\sfrac{x}{g}){\mathcal N}(x)Y(x)}{H(x)},
\end{array}
\end{equation}
where $f, g, c_0$ and $c_1$ are constants with respect to $x$.

The contiguity relations:
\begin{equation}\label{eq:E6L2L3}
\begin{array}{l}
L_2(x): C_0x(1-xf)\o{y}(x)-(1-a_2x)y(qx)+(1-\sfrac{x}{g})y(x)=0,\\
L_3(x): C_1x(1-\sfrac{x\o{f}}{q})y(x)+(1-a_1x)(1-\sfrac{x}{q^{m+n}})(1-\sfrac{x}{qg})\o{y}(x)\\
\phantom{L_3(x): C_1x(1-\sfrac{x\o{f}}{q})y(x)}-(1-x)(1-\sfrac{b_1x}{q})(1-\sfrac{b_2x}{q})\o{y}(\sfrac{x}{q})=0,
\end{array}
\end{equation}
where $C_0=\sfrac{c_0}{c_1}$ and $C_1=\sfrac{\o{c}_0(1-a_1)}{qc_1}$.\\

\noindent
{\bf(c)} The $q$-Painlev\'e equation\\

Compatibility gives the following equations:
\begin{equation}\label{eq:E6eq}
\begin{array}{l}
\dfrac{(fg-1)(f\u{g}-1)}{g\u{g}}=\dfrac{(f-a_2)(f-b_1)(f-b_2)(f-q)}{(f-a_1)(f-\sfrac{1}{q^{m+n}})},
\\
\dfrac{(fg-1)(\o{f}g-1)}{f\o{f}}=\dfrac{(g-\sfrac{1}{a_2})(g-\sfrac{1}{b_1})(g-\sfrac{1}{b_2})(g-\sfrac{1}{q})}{(g-\sfrac{1}{a_2q^m})(g-\sfrac{a_1}{b_1b_2q^n})}.
\end{array}
\end{equation}
These equations (\ref{eq:E6eq}) are equivalent to the $q$-Painlev\'e equation of type $E_6^{(1)}$ given in \cite{Ikawa13} \cite{KMNOY04} \cite{RGTT01}. 
The 8 singular points are on the two lines $f=0$ and $g=0$ and one curve $fg=1$.
\begin{equation}
\begin{array}{l}
(f,g)=(a_2,\sfrac{1}{a_2}),(b_1,\sfrac{1}{b_1}),(b_2,\sfrac{1}{b_2}),(q,\sfrac{1}{q}),\\
\phantom{(f,g)=}(a_1,0),(\sfrac{1}{q^{m+n}},0),(0,\sfrac{1}{a_2q^m}),(0,\sfrac{a_1}{b_1b_2q^n}).
\end{array}
\end{equation}

The product $C_0C_1$:
\begin{equation}\label{eq:E6C0C1}
C_0C_1=\dfrac{(1-a_2g)(1-b_1g)(1-b_2g)(1-qg)}{qg^2(1-fg)(1-\o{f}g)}.
\end{equation}\\

\noindent
{\bf(d)} The Lax pair
\begin{equation}\label{eq:E6L1L2}
\begin{array}{l}
L_1(x):\dfrac{q^{m+n}g(1-x)(q-b_1x)(q-b_2x)}{q-fx}\left[y(\sfrac{x}{q})-\dfrac{g(q-a_2x)}{qg-x}y(x)\right] 
\\[5mm]
\phantom{L_1(x):}+\dfrac{g(q^{m+n}-x)(1-a_1x)(1-a_2x)}{1-fx} \left[y(qx)-\dfrac{(g-x)}{g(1-a_2x)}y(x)\right] 
\\[5mm]
\phantom{L_1(x):}+x^2\left[\dfrac{(a_2q^mg-1) (b_1b_2q^ng-a_1)}{f}-\dfrac{q^{m+n}(a_2g-1)(b_1g-1)(b_2g-1)(qg-1)}{(fg-1)(qg-x)}\right]y(x)=0, 
\\[5mm]
L_2(x): C_0x(1-xf)\o{y}(x)-(1-a_2x)y(qx)+(1-\sfrac{x}{g})y(x)=0.
\end{array}
\end{equation}
The scalar Lax pair (\ref{eq:E6L1L2}) is equivalent to the 2 $\times$ 2 matrix ones
in \cite{Sakai06} \cite{WO12} and the scalar ones in \cite{Ikawa13} \cite{Yamada11} using suitable gauge transformations of $y(x)$. (Note that there are some typographical errors in equations (30) and (31) in \cite{Ikawa13}, namely the expressions $(b_4q^{\prime}-z)$ and $Y(x)$ should read $(b_4q^{\prime}-z)t^2$ and $Y(z)$ respectively.)\\

\noindent
{\bf(e)} Special solutions
\begin{equation}\label{eq:E6sol}
\begin{array}{l}
\dfrac{1-\sfrac{f}{a_1}}{1-\sfrac{f}{a_2}}=\dfrac{\gamma_1 T_{a_1}(\tau_{m,n})T_{a_1}^{-1}(\tau_{m+1,n-1})}{\gamma_2 T_{a_2}(\tau_{m,n})T_{a_2}^{-1}(\tau_{m+1,n-1})},\quad
\dfrac{1-\sfrac{1}{b_1g}}{1-\sfrac{1}{b_2g}}=\dfrac{\omega_1 T_{b_1}^{-1}(\tau_{m,n})T_{b_1}(\o{\tau}_{m+1,n-1})}{\omega_2 T_{b_2}^{-1}(\tau_{m,n})T_{b_2}(\o{\tau}_{m+1,n-1})},
\end{array}
\end{equation}
where 
\begin{equation}\label{eq:E6tau}
\tau_{m,n}=\det\Big[{}_3\varphi_2 \Big(\substack{\displaystyle{b_1,b_2,q^{-(m+n)}}\\[3mm]{\displaystyle{a_1,a_2}}};q,q^{i+j+1}\Big)\Big]^n _{i,j=0},
\end{equation}
\begin{equation}\label{eq:E6gamma}
\begin{array}{l}
\gamma_i=\dfrac{a_i(1-a_iq^{m+n})(1-\sfrac{a_i}{q})^n\prod_{k=1}^{2}(1-\sfrac{b_k}{a_i})}{(1-a_i)^{n+1}},
\quad \omega_i=\dfrac{(1-\sfrac{a_2}{b_i})(1-b_i)^n}{(1-\sfrac{b_i}{q})^n},\quad \mbox{for $i=1, 2$}.
\end{array}
\end{equation}
These determinant formulae of hypergeometric solutions (\ref{eq:E6sol}) are expressed in terms of the terminating ${}_3\varphi_2$ (\ref{eq:qHGF}) series (big $q$-Jacobi polynomials \cite{KMNOY04}), and they are equivalent to that in \cite{Ikawa13}. (Note that there is a typographical error in equation (38) in \cite{Ikawa13}, namely the expression $T_{a_2}T_{a_3}(\tau_{m,n-1})$ should read $T_{a_2}T_{a_4}(\tau_{m,n-1})$.)\\

\subsection{Case $q$-$D_5^{(1)}$}\label{subsec:D5}　\\

\noindent
{\bf(a)} Setting of the Pad\'e interpolation problem\\

The interpolated function and the interpolated sequence (\ref{eq:Ylist}):
\begin{equation}\label{eq:D5Y}
Y(x)=c^{\log_q x}\dfrac{(a_1 x, b_1;q)_\infty}{(a_1, b_1 x;q)_\infty},\quad Y_s=c^s\dfrac{(b_1;q)_s}{(a_1;q)_s}.
\end{equation}

The time evolution (\ref{eq:Tlist}):
\begin{equation}\label{eq:D5T}
T: (a_1,b_1,c,m,n) \mapsto (qa_1,b_1,c,m-1,n).
\end{equation}\\

\noindent
{\bf(b)} Contiguity relations\\

The basic quantities:
\begin{equation}\label{eq:D5GKH}
G(x)=\dfrac{(1-b_1x)c}{1-a_1x},\quad K(x)=\dfrac{1-a_1}{1-a_1x},\quad H(x)=1-a_1x.
\end{equation}

The Casorati determinants:
\begin{equation}\label{eq:D5D}
\begin{array}{l}
D_1(x)=\dfrac{c_0(1-xf){\mathcal N}(x)Y(x)}{G_{\rm den}(x)},\quad
D_2(x)=\dfrac{c_1N(x)Y(x)}{K_{\rm den}(x)},\quad D_3(x)=\dfrac{c_2N(x)Y(x)}{H(x)},
\end{array}
\end{equation}
where $f, c_0, c_1$ and $c_2$ are constants with respect to $x$.

\noindent
The contiguity relations:
\begin{equation}\label{eq:D5L2L3}
\begin{array}{l}
L_2(x): C_0(1-xf)\o{y}(x)-y(qx)+\sfrac{y(x)}{g}=0,\\
L_3(x): C_1(1-\sfrac{x\o{f}}{q})y(x)-(1-a_1x)(1-\sfrac{x}{q^{m+n}})\o{y}(x)/g
\\
\phantom{L_3(x): C_1(1-\sfrac{x\o{f}}{q})y(x)}+c(1-x)(1-\sfrac{b_1x}{q})\o{y}(\sfrac{x}{q})=0,
\end{array}
\end{equation}
where $C_0=\sfrac{c_0}{c_1}, C_1=-\sfrac{\o{c}_0(1-a_1)}{c_1}$ and $g=\sfrac{c_1}{c_2}$.\\

\noindent
{\bf(c)} The $q$-Painlev\'e equation\\

Compatibility gives the following equations:
\begin{equation}\label{eq:D5eq}
g\u{g}=\dfrac{(f-a_1)(f-\sfrac{1}{q^{m+n}})}{c(f-b_1)(f-q)},\quad
f\o{f}=\dfrac{qb_1(g-\sfrac{1}{q^m})(g-\sfrac{a_1}{b_1q^nc})}{(g-1)(g-\sfrac{1}{c})}.
\end{equation}
These equations (\ref{eq:D5eq}) are equivalent to the $q$-Painlev\'e equation of type $D_5^{(1)}$ given in \cite{JS96} \cite{KMNOY04}. 
The 8 singular points are on the four lines $f=0$, $f=\infty$, $g=0$ and $g=\infty$.
\begin{equation}
\begin{array}{l}
(f,g)=(a_1,0), (\sfrac{1}{q^{m+n}},0), (0,\sfrac{1}{q^m}), (0,\sfrac{a_1}{b_1cq^n}),
(\infty,1), (\infty,\sfrac{1}{c}), (b_1,\infty), (q,\infty).
\end{array}
\end{equation}

The product $C_0C_1$:
\begin{equation}\label{eq:D5C0C1}
C_0C_1=-\sfrac{(1-g)(1-cg)}{g^2}.
\end{equation}\\
\noindent
{\bf(d)} The Lax pair
\begin{equation}\label{eq:D5L1L2}
\begin{array}{l}
L_1(x):\dfrac{g(x-q^{m + n})(a_1x-1)}{fx-1}\left[y(qx)-\dfrac{y(x)}{g}\right]
+\dfrac{cq^{m + n}g(x-1) (b_1x-q)}{fx-q}\left[y(\dfrac{x}{q})-gy(x)\right] \\
\phantom{L_1(x):}+\left[\dfrac{(q^mg-1 ) (b_1cq^ng-a_1)x}{f}-q^{m+n}(g-1) (cg-1)\right] y(x)
=0,
\\
L_2(x): C_0(1-xf)\o{y}(x)-y(qx)+\sfrac{y(x)}{g}=0.
\end{array}
\end{equation}
The scalar Lax pair (\ref{eq:D5L1L2}) is equivalent to the 2 $\times$ 2 matrix ones in \cite{JS96} \cite{Murata09} and the scalar one in \cite{Yamada11} using suitable gauge transformations of $y(x)$.\\

\noindent
{\bf(e)} Special solutions
\begin{equation}\label{eq:D5sol}
\begin{array}{l}
\dfrac{f}{a_1}-1=\dfrac{c(1-\sfrac{b_1}{a_1})(1-a_1q^{m+n})(1-\sfrac{a_1}{q})^n}{q^m(1-c)(1-a_1)^{n+1}}\dfrac{T_{a_1}(\tau_{m,n,1})T_{a_1}^{-1}(\tau_{m+1,n-1,1})}{\tau_{m,n,0}\tau_{m+1,n-1,2}},
\\
g=\dfrac{(1-a_1q^{m+n})(1-\sfrac{b_1}{q})^n}{q^m(1-a_1)(1-b_1)^n}
\dfrac{T_{a_1}(\tau_{m,n,1})\tau_{m,n-1,1}}{T_{b_1}^{-1}(\tau_{m,n,1})T_{a_1b_1}(\tau_{m,n-1,1})},
\end{array}
\end{equation}
where 
\begin{equation}\label{eq:D5tau}
\tau_{m,n,k}=\det\Big[{}_2\varphi_1 \Big(\substack{\displaystyle{b_1,q^{-(m+n)}}\\[3mm]{\displaystyle{a_1}}};q,cq^{i+j+k}\Big)\Big]^n _{i,j=0}.
\end{equation}
These determinant formulae of hypergeometric solutions (\ref{eq:D5sol}) are expressed in terms of the terminating ${}_2\varphi_1$ (\ref{eq:qHGF}) series (little $q$-Jacobi polynomials \cite{KMNOY04}), and they are expected to be equivalent to the terminating case of that in \cite{Sakai98}.\\
\subsection{Case $q$-$A_4^{(1)}$}\label{subsec:A4}　\\

\noindent
{\bf(a)} Setting of the Pad\'e interpolation problem\\

The interpolated function and the interpolated sequence (\ref{eq:Ylist}):
\begin{equation}\label{eq:A4Y}
Y(x)=c^{\log_q x}\dfrac{(b_1;q)_\infty}{(b_1 x;q)_\infty},\quad Y_s=c^s(b_1;q)_s.
\end{equation}

The time evolution (\ref{eq:Tlist}):
\begin{equation}\label{eq:A4T}
T: (b_1,c,m,n) \mapsto (b_1,c,m-1,n).
\end{equation}\\

\noindent
{\bf(b)} Contiguity relations\\

The basic quantities:
\begin{equation}\label{eq:A4GKH}
G(x)=(1-b_1x)c,\quad K(x)=1,\quad H(x)=1.
\end{equation}
The Casorati determinants:
\begin{equation}\label{eq:A4}
\begin{array}{l}
D_1(x)=\dfrac{c_0(1-xf){\mathcal N}(x)Y(x)}{G_{\rm den}(x)},\quad
D_2(x)=\dfrac{c_1N(x)Y(x)}{K_{\rm den}(x)},\quad D_3(x)=\dfrac{c_2N(x)Y(x)}{H(x)},
\end{array}
\end{equation}
where $f, c_0, c_1$ and $c_2$ are constants with respect to $x$.

The contiguity relations:
\begin{equation}\label{eq:A4L2L3}
\begin{array}{l}
L_2(x): C_0(1-xf)\o{y}(x)-y(qx)+\sfrac{y(x)}{g}=0,
\\
L_3(x): C_1(1-x\o{f}/q)y(x)-\sfrac{(1-\sfrac{x}{q^{m+n}})\o{y}(x)}{g}+c(1-x)(1-b_1x/q)\o{y}(x/q)=0,
\end{array}
\end{equation}
where $C_0=\sfrac{c_0}{c_1}$, $C_1=-\sfrac{\o{c}_0}{c_1}$ and $g=\sfrac{c_1}{c_2}$.\\

\noindent
{\bf(c)} The $q$-Painlev\'e equation\\

Compatibility gives the following equations:
\begin{equation}\label{eq:A4eq}
g\u{g}=\dfrac{f(f-\sfrac{1}{q^{m+n}})}{c(f-b_1)(f-q)},\quad
f\o{f}=\dfrac{qb_1g(g-\sfrac{1}{q^m})}{(g-1)(g-\sfrac{1}{c})}.
\end{equation}
These equations (\ref{eq:A4eq}) are equivalent to the $q$-Painlev\'e equation of type $A_4^{(1)}$ given in \cite{KMNOY04} \cite{KTGR00}. 
The 8 singular points are on the four lines $f=0, f=\infty, g=0$ and $g=\infty$. $(0,0)$ is a double point.
\begin{equation}
\begin{array}{l}
(f,g)=(\sfrac{1}{q^{m+n}},0), (0,\sfrac{1}{q^m}), (\infty,1), (\infty,\sfrac{1}{c}),
(b_1,\infty), (q,\infty),\\
\phantom{(f,g)=}(0,0)\leftarrow \sfrac{g}{f}=-\sfrac{1}{b_1cq^n}.
\end{array}
\end{equation}

The product $C_0C_1$:
\begin{equation}\label{eq:A4C0C1}
C_0C_1=-\sfrac{(1-g)(1-cg)}{g^2}.
\end{equation}\\

\noindent
{\bf(d)} The Lax pair
\begin{equation}\label{eq:A4L1L2}
\begin{array}{l}
L_1(x):\dfrac{c g q^{m + n} (1 - x) (q - b_1 x)}{q - f x}\left[y(\sfrac{x}{q})-gy(x)\right]+\dfrac{g(q^{m + n} - x)}{1-f x}\left[y(q x)-\sfrac{y(x)}{g}\right]
 \\
\phantom{L_1(x):}+q^n\left[q^m(1-g) (1-cg) + \sfrac{b_1cg (1-q^mg) x}{f}\right]y(x)=0,\\
L_2(x): C_0(1-xf)\o{y}(x)-y(qx)+\sfrac{y(x)}{g}=0.
\end{array}
\end{equation}
The scalar Lax pair (\ref{eq:A4L1L2}) is equivalent to the 2 $\times$ 2 matrix one for the $q$-Painlev\'e equation of type $q$-$P(A_4)$ in \cite{Murata09} using a suitable gauge transformation of $y(x)$.\\

\noindent
{\bf(e)} Computation of special solutions
\begin{equation}\label{eq:A4sol}
\begin{array}{l}
f=\dfrac{b_1c}{(c-1)q^m}
\dfrac{\tau_{m, n, 1}\tau_{m+1, n-1, 1}}{\tau_{m, n, 0}\tau_{m+1, n-1, 2}},\quad g=\dfrac{(1-\sfrac{b_1}{q})^{n}}{q^m(1-b_1)^n}\dfrac{\tau_{m, n, 1}\tau_{m, n-1, 1}}{T_{b_1}^{-1}(\tau_{m, n, 1})T_{b_1}(\tau_{m, n-1, 1})},
\end{array}
\end{equation}
where 
\begin{equation}\label{eq:A4tau}
\tau_{m, n, k}=\det\Big[{}_2\varphi_1 \Big(\substack{\displaystyle{b_1,q^{-(m+n)}}\\[3mm]{\displaystyle{0}}};q,cq^{i+j+k}\Big)\Big]^n _{i,j=0}.
\end{equation}
These determinant formulae of hypergeometric solutions (\ref{eq:A4sol}) are expressed in terms of the terminating ${}_2\varphi_1$ (\ref{eq:qHGF}) series ($q$-Laguerre polynomials
 \cite{KMNOY04}), and they are expected to be equivalent to the terminating case of that in \cite{HK07}.\\
\subsection{Case $q$-$(A_2+A_1)^{(1)}$}\label{subsec:qA21}　\\

\noindent
{\bf(a)} Setting of the Pad\'e interpolation problem\\

The interpolated function and the interpolated sequence (\ref{eq:Ylist}):
\begin{equation}\label{eq:A21Y}
Y(x)=(d\sqrt{\sfrac{x}{q}})^{\log_q x},\quad Y_s=q^{\left(\substack{s\\2}\right)}d^s. 
\end{equation}

The time evolution (\ref{eq:Tlist}):
\begin{equation}\label{eq:A21T}
T: (d,m,n) \mapsto (d,m-1,n).
\end{equation}\\

\noindent
{\bf(b)} Contiguity relations\\

The basic quantities:
\begin{equation}\label{eq:A21GKH}
G(x)=dx,\quad K(x)=1,\quad H(x)=1.
\end{equation}

The Casorati determinants:
\begin{equation}\label{eq:A21D}
\begin{array}{l}
D_1(x)=\dfrac{c_0(1-xf){\mathcal N}(x)Y(x)}{G_{\rm den}(x)},\quad
D_2(x)=\dfrac{c_1N(x)Y(x)}{K_{\rm den}(x)},\quad D_3(x)=\dfrac{c_2N(x)Y(x)}{H(x)},
\end{array}
\end{equation}
where $f, c_0, c_1$ and $c_2$ are constants with respect to $x$.

The contiguity relations:
\begin{equation}\label{eq:A21L2L3}
\begin{array}{l}
L_2(x): C_0(1-xf)\o{y}(x)-y(qx)+\sfrac{y(x)}{g}=0,\\
L_3(x): C_1(1-x\o{f}/q)y(x)+\sfrac{(1-\sfrac{x}{q^{m+n}})\o{y}(x)}{g}-\sfrac{d x(1-x)\o{y}(x/q)}{q}=0,
\end{array}
\end{equation}
where $C_0=\sfrac{c_0}{c_1}$, $C_1=\sfrac{\o{c}_0}{c_1}$ and $g=\sfrac{c_1}{c_2}$.\\

\noindent
{\bf(c)} The $q$-Painlev\'e equation\\

Compatibility gives the following equation:
\begin{equation}\label{eq:A21eq}
g\u{g}=\dfrac{f(f-\sfrac{1}{q^{m+n}})}{d(f-q)},\quad
f\o{f}=\dfrac{qdg(g-\sfrac{1}{q^m})}{(g-1)}.
\end{equation}
These equations (\ref{eq:A21eq}) are equivalent to the $q$-Painlev\'e equation of type $(A_2+A_1)^{(1)}$, namely $P_{\rm III}$, given in \cite{KMNOY04} \cite{KTGR00}. 
The 8 singular points are on the four lines $f=0, f=\infty, g=0$ and $g=\infty$. The two points $(0,0)$ and $(\infty, \infty)$ are double points.
\begin{equation}
\begin{array}{l}
(f,g)=(\sfrac{1}{q^{m+n}},0), (0,\sfrac{1}{q^m}), (\infty,1), 
(q,\infty),\\
 \phantom{(f,g)=}(0,0)\leftarrow \sfrac{g}{f}=\sfrac{1}{dq^n},\\
 \phantom{(f,g)=}(\infty,\infty)\leftarrow \sfrac{g}{f}=\sfrac{1}{d}.
\end{array}
\end{equation}

The product $C_0C_1$:
\begin{equation}\label{eq:A21C0C1}
C_0C_1=\sfrac{(1-g)}{g^2}.
\end{equation}\\

\noindent
{\bf(d)} The Lax pair
\begin{equation}\label{eq:A21L1L2}
\begin{array}{l}
L_1(x):\dfrac{d g q^{m + n} (1-x) x }{
  f x-q}\left[y(\sfrac{x}{q})-gy(x)\right] + \dfrac{g(q^{m + n} - x)}{f x-1} \left[y(q x)-\sfrac{y(x)}{g}\right]\\

\phantom{L_1(x):}
  +q^n\left[q^m(g-1) - \sfrac{d g (g q^m-1)x}{f}\right]y(x)=0,\\
  L_2(x): C_0(1-xf)\o{y}(x)-y(qx)+\sfrac{y(x)}{g}=0.
\end{array}
\end{equation}
The scalar Lax pair (\ref{eq:A21L1L2}) is equivalent to the 2 $\times$ 2 matrix one for the $q$-Painlev\'e equation of type $q$-$P(A_5)^{\#}$ in \cite{Murata09} using a suitable gauge transformation of $y(x)$.\\

\noindent
{\bf(e)} Special solutions
\begin{equation}\label{eq:A21sol}
\begin{array}{l}
f=\dfrac{d}{q^m}\dfrac{\tau_{m, n, 1}\tau_{m+1, n-1, 1}}{\tau_{m, n, 0}\tau_{m+1, n-1, 2}},\quad g=
\dfrac{1}{q^{m+n}}\dfrac{\tau_{m, n, 1}\tau_{m, n-1, 1}}{\tau_{m, n, 0}\tau_{m, n-1, 2}},
\end{array}
\end{equation}
where 
\begin{equation}\label{eq:A21tau}
\tau_{m, n, k}=\det\Big[{}_1\varphi_1 \Big(\substack{\displaystyle{q^{-(m+n)}}\\[3mm]{\displaystyle{0}}};q,-dq^{i+j+k}\Big)\Big]^n _{i,j=0}.
\end{equation}
These determinant formulae of hypergeometric solutions (\ref{eq:A21sol}) are expressed in terms of the terminating ${}_1\varphi_1$ (\ref{eq:qHGF}) series (Stieltjes--Wigert polynomials \cite{KMNOY04}), and they are expected to be  equivalent to the terminating case of that in \cite{Nakazono10}.\\
\section{Conclusion}\label{sec:conc}
\subsection{Summary}　\\

\noindent
In this paper, for the {\it interpolated function} $Y(x)$ given in Table (\ref{eq:Ylist}), we established the {\it Pad\'e interpolation problem} related to all the $q$-Painlev\'e equations of types from $E_7^{(1)}$ to $(A_2+A_1)^{(1)}$. Then, for the {\it time evolution} $T$ given in Table (\ref{eq:Tlist}), we established another Pad\'e interpolation problem. By solving these problems, we derived the evolution equations, the scalar Lax pairs and the determinant formulae of the special solutions for the corresponding $q$-Painlev\'e equations. The main results are given in section \ref{sec:interpolation results}.

\subsection{Problems}　\\

\noindent
Some open problems related to the results of this paper are as follows:

{\bf 1.} In this paper,  by choosing one time evolution $T$, we applied the Pad\'e method to each type from $q$-$E_7^{(1)}$ to $q$-$(A_2+A_1)^{(1)}$. By choosing other time evolutions, we can perform similar computations. It may be interesting to study 
the relation between the Pad\'e method for the various time evolutions and Ba\"cklund transformations of the affine Weyl group, for example the case $q$-$E_6^{(1)}$ in \cite{Ikawa13}.

{\bf 2.} By the results of this paper and previous ones \cite{Ikawa13}\cite{NTY13}\cite{Yamada14}, it turned out that the Pad\'e method could be applied to all the discrete Painlev\'e equations of types from $e$-$E_8^{(1)}$ to $q$-$(A_2+A_1)^{(1)}$. It may be interesting to study the degenerations between these results.

{\bf 3.} In this paper, we applied the Pad\'e method of the $q$-grid to the $q$-Painlev\'e equation of type $D_5^{(1)}$. On the other hand, the method of the differential grid (i.e., Pad\'e approximation) was also applied to the $q$-$P_{\rm VI}$ equation in \cite{Ikawa13}\cite{Yoshioka10}. It may be interesting to study whether the method of the differential grid can be also applied to the $q$-Painlev\'e equations of other types.

{\bf 4.} It may be interesting to study whether the Pad\'e method can be further applied to the other generalized Painlev\'e systems, for example the $q$-Garnier system in \cite{Sakai05}.\\

\section*{Acknowledgment}

\noindent
The author is grateful to Professor Y. Yamada for valuable discussions on this research, and for his encouragement. The author also thanks Professors T. Suzuki, T. Tsuda, H. Sakai, M. Noumi, W. Rossman and the referee for stimulating comments and for kindhearted support. 


\end{document}